\documentclass[12pt]{article}
\usepackage{amssymb,amsfonts}
\usepackage{amsmath}
\usepackage{epsf,epsfig}
\begin{document}
\baselineskip 18pt

\title{$R$-matrix for the XX spin chain in a staggered magnetic field}
\author{P.~N.~Bibikov}
\date{\it Russian State Hydrometeorological University, Saint-Petersburg, Russia}

\maketitle

\vskip5mm

\begin{abstract}
We suggest the elliptic $16\times16$ $R$-matrix for the $XX$ spin chain in a staggered magnetic field. This integrable model is a special case of the more general one, studied long ago within the alternative approach by V. M. Kontorovich and V. M. Tsukernik. The Yang-Baxter equation is verified with the full rigorous only in the two special degenerate trigonometric cases. For the general one, studying the first terms of the Taylor expansions for the $R$-matrix entries, we give the strong evidence for implementation of the Yang-Baxter equation.
\end{abstract}

\maketitle
\vskip20mm
\section{Introduction}

The XX model, suggested in \cite{1}, and related to the Hamiltonian
\begin{equation}\label{hamgen}
\hat H=\sum_nH_{n,n+1},
\end{equation}
where
\begin{equation}\label{loc0}
H_{n,n+1}=\frac{1}{2}\Big({\bf S}^+_n{\bf S}^-_{n+1}+{\bf S}^-_n{\bf S}^+_{n+1}\Big),
\end{equation}
is one of the most studied integrable spin chains (see \cite{2} and references therein). Here ${\bf S}^j_n$ ($j=+,-,z$) are the spin-1/2 operators
associated with the $n$-th site of the chain.

The integrability of \eqref{hamgen}, \eqref{loc0} has been first demonstrated within the Jordan-Wigner transformation to the free-fermion model
\cite{1} and then was confirmed by various alternative approaches \cite{2}. Namely, the Hamiltonian density $4\times4$ matrix
\begin{equation}\label{dens0}
H=\frac{1}{2}\Big({\bf S}^+\otimes{\bf S}^-+{\bf S}^-\otimes{\bf S}^+\Big),
\end{equation}
related to \eqref{loc0}, may be represented in the form
\begin{equation}\label{hr}
H=\frac{1}{2}\frac{dR(\lambda)}{d\lambda}\Big|_{\lambda=0},
\end{equation}
where the $R$-matrix
\begin{equation}
R(\lambda)=\left(\begin{array}{cccc}
\cos{\lambda}&0&0&0\\
0&1&\sin{\lambda}&0\\
0&\sin{\lambda}&1&0\\
0&0&0&\cos{\lambda}
\end{array}\right),
\end{equation}
satisfies the Yang-Baxter equation (in the Braid group form) \cite{4}
\begin{equation}\label{YB}
R_{12}(\lambda-\mu)R_{23}(\lambda)R_{12}(\mu)=R_{23}(\mu)R_{12}(\lambda)R_{23}(\lambda-\mu),
\end{equation}
(as usual, $R_{12}(\lambda)\equiv R(\lambda)\otimes I_2$ and $R_{23}(\lambda)\equiv I_2\otimes R(\lambda)$) and the initial condition
\begin{equation}\label{init}
R(0)=I_4.
\end{equation}
Here by $I_m$ we denote the $m\times m$ identity matrix.

Within the generalized version of the \cite{1} approach, it has been proved \cite{3}, that the model \eqref{hamgen}, \eqref{loc0} remains integrable even under addition of a staggered longitudinal magnetic field, when \eqref{loc0} is replaced by
\begin{equation}\label{loc}
H_{n,n+1}(h_{\rm st})=\frac{1}{2}\Big({\bf S}^+_n{\bf S}^-_{n+1}+{\bf S}^-_n{\bf S}^+_{n+1}+h_{\rm st}[(-1)^n{\bf S}^z_n+(-1)^{n+1}{\bf S}^z_{n+1}]\Big).
\end{equation}

It is desirable to confirm this result by presentation of the corresponding $R$-matrix.
Since, however, \eqref{loc} is invariant only under even-steps translations ($H_{n,n+1}=H_{n+2,n+3}\neq H_{n+1,n+2}$), the approach based on a $4\times4$ Hamiltonian density matrix needs a modification.
Namely, \eqref{hamgen} should be replaced by
\begin{equation}\label{hamgenm}
\hat H(h_{\rm st})=\sum_n{\cal H}_{2n,2n+3},
\end{equation}
where the $16\times16$ matrix ${\cal H}$ has the form
\begin{equation}\label{Hst}
{\cal H}=\frac{1}{2}\Big(H^{(+)}\otimes I_4+I_4\otimes H^{(+)}\Big)+
I_2\otimes H^{(-)}\otimes I_2,
\end{equation}
and
\begin{equation}
H^{(\pm)}=\frac{1}{2}\Big[{\bf S}^+\otimes{\bf S}^-+{\bf S}^-\otimes{\bf S}^+\pm h_{\rm st}
\Big({\bf S}^z\otimes I_2-I_2\otimes{\bf S}^z\Big)\Big].
\end{equation}

The direct calculation shows, that the matrix \eqref{Hst} satisfies the Reshetikhin condition \cite{4}
\begin{equation}\label{Resh}
[{\cal H}\otimes I_4+I_4\otimes{\cal H},[{\cal H}\otimes I_4,I_4\otimes{\cal H}]]=I_4\otimes{\cal K}-{\cal K}\otimes I_4,
\end{equation}
(the $16\times16$ matrix ${\cal K}$ may be readily calculated), which is the strong evidence for an existence of the the corresponding
$16\times16$ $R$-matrix ${\cal R}(\lambda,h_{\rm st})$.

Since the entries of all {\it known} $R$-matrices depend only on rational, trigonometric (hyperbolic) or elliptic functions, it is natural to suppose,
that this also should be implemented for ${\cal R}(\lambda,h_{\rm st})$.

Under this assumption, one may try to guess the entries of ${\cal R}(\lambda,h_{\rm st})$, analyzing the first terms of the series expansions for their ratios. The corresponding method has been first suggested
in \cite{5}, and then successfully used in \cite{6,7}. In the next Section we present ${\cal R}(\lambda,h_{\rm st})$, obtained with the use of the developed version of the approach \cite{5} (the details of calculations will be published elsewhere).

\section{The $R$-matrix}

The suggested $R$-matrix is symmetric
\begin{equation}\label{rels1}
{\cal R}_{ji}(\lambda,h_{\rm st})={\cal R}_{ij}(\lambda,h_{\rm st}),
\end{equation}
and has the following structure
\begin{equation}\label{template}
{\cal R}(\lambda,h_{\rm st})=\left(\begin{array}{cccccccccccccccc}
*&0&0&0&0&0&0&0&0&0&0&0&0&0&0&0\\
0&*&*&0&*&0&0&0&0&0&0&0&0&0&0&0\\
0&*&*&0&*&0&0&0&*&0&0&0&0&0&0&0\\
0&0&0&*&0&*&*&0&0&*&*&0&*&0&0&0\\
0&*&*&0&*&0&0&0&*&0&0&0&0&0&0&0\\
0&0&0&*&0&*&*&0&0&*&*&0&*&0&0&0\\
0&0&0&*&0&*&*&0&0&*&*&0&*&0&0&0\\
0&0&0&0&0&0&0&*&0&0&0&*&0&*&0&0\\
0&0&*&0&*&0&0&0&0&0&0&0&0&0&0&0\\
0&0&0&*&0&*&*&0&0&*&*&0&*&0&0&0\\
0&0&0&*&0&*&*&0&0&*&*&0&*&0&0&0\\
0&0&0&0&0&0&0&*&0&0&0&*&0&*&*&0\\
0&0&0&*&0&*&*&0&0&*&*&0&*&0&0&0\\
0&0&0&0&0&0&0&*&0&0&0&*&0&*&*&0\\
0&0&0&0&0&0&0&0&0&0&0&*&0&*&*&0\\
0&0&0&0&0&0&0&0&0&0&0&0&0&0&0&*\\
\end{array}\right).
\end{equation}
In order to represent the matrix elements in the more compact form, it is convenient to introduce the following variables
\begin{equation}\label{ABdef}
{\cal A}_{ij}=\frac{{\cal R}_{ij}(\lambda,h_{\rm st})+{\cal R}_{17-i,17-j}(\lambda,h_{\rm st})}{2},\qquad
{\cal B}_{ij}=\frac{{\cal R}_{ij}(\lambda,h_{\rm st})-{\cal R}_{17-i,17-j}(\lambda,h_{\rm st})}{2},
\end{equation}
for which
\begin{equation}\label{ABgenrel}
{\cal A}_{ji}={\cal A}_{ij},\quad{\cal B}_{ji}={\cal B}_{ij},\quad{\cal A}_{17-i,17-j}={\cal A}_{ij},\quad{\cal B}_{17-i,17-j}=-{\cal B}_{ij}.
\end{equation}

Modulo \eqref{ABgenrel} and \eqref{template}, one has
\begin{eqnarray}
&&{\cal A}_{22}={\cal A}_{88}=\theta{\cal A}_{11},\qquad{\cal A}_{33}={\cal A}_{55}=(1+2\alpha^2+4h_{\rm st}\alpha\gamma){\cal A}_{22},\nonumber\\
&&{\cal A}_{44}=(1+\alpha^2)(1+\alpha^2+2h_{\rm st}\alpha\gamma){\cal A}_{11},\qquad{\cal A}_{66}=[1+(3+8h_{\rm st}^2)\alpha^2+\alpha^4]{\cal A}_{11},
\nonumber\\
&&{\cal A}_{77}=(1+\alpha^2){\cal A}_{11},\qquad{\cal A}_{23}={\cal A}_{59}=\alpha{\cal A}_{22},\qquad
{\cal A}_{25}={\cal A}_{39}={\cal A}_{7,10}=\alpha^2{\cal A}_{11},\nonumber\\
&&{\cal A}_{46}=\alpha(2+\alpha^2+4h_{\rm st}\alpha\gamma){\cal A}_{22},\qquad{\cal A}_{47}={\cal A}_{4,10}=\alpha^2{\cal A}_{22},\qquad
{\cal A}_{4,11}=\alpha^3{\cal A}_{22},\nonumber\\
&&{\cal A}_{4,13}=\alpha^4{\cal A}_{11},\qquad{\cal A}_{67}={\cal A}_{6,10}=
\alpha(1+\alpha^2+2h_{\rm st}\alpha\gamma){\cal A}_{11},\nonumber\\
&&{\cal A}_{35}=2{\cal A}_{67},\qquad{\cal A}_{6,11}=\alpha^2(1+\alpha^2){\cal A}_{11},
\end{eqnarray}
and
\begin{eqnarray}
&&{\cal B}_{11}={\cal B}_{77}={\cal B}_{25}={\cal B}_{35}={\cal B}_{39}=0,\qquad{\cal B}_{22}={\cal B}_{88}=\gamma{\cal A}_{22},\nonumber\\
&&{\cal B}_{55}=\gamma(3+2\alpha^2+4h_{\rm st}\alpha\gamma){\cal A}_{22},
\qquad{\cal B}_{33}=-{\cal B}_{55},\nonumber\\
&&{\cal B}_{44}=2h_{\rm st}\alpha(\alpha^2-1){\cal A}_{11},\qquad{\cal B}_{46}=\alpha\gamma(2+3\alpha^2+4h_{\rm st}\alpha\gamma){\cal A}_{22},\nonumber\\
&&{\cal B}_{47}={\cal B}_{4,10}=\alpha^2\gamma{\cal A}_{22},\qquad{\cal B}_{59}=\alpha\gamma{\cal A}_{22},
\qquad{\cal B}_{23}=-{\cal B}_{59},\nonumber\\
&&{\cal B}_{67}={\cal B}_{6,10}=2h_{\rm st}\alpha^2{\cal A}_{11},\qquad
{\cal B}_{4,11}=-\alpha^3\gamma{\cal A}_{22},\qquad{\cal B}_{66}=4h_{\rm st}{\cal A}_{67}.
\end{eqnarray}
Here the three elliptic functions $\alpha=\alpha(\lambda)$, $\gamma=\gamma(\lambda)$ and $\theta=\theta(\lambda)$ are
\begin{equation}
\alpha\equiv i\kappa{\rm sn}(\lambda/(i\kappa),\kappa^2),\qquad
\gamma\equiv\xi{\rm sn}(\zeta\lambda,\xi^2),\qquad
\theta\equiv{\rm cn}(i\eta\lambda,1/\eta),
\end{equation}
where
\begin{equation}
\kappa\equiv\sqrt{1+h^2_{\rm st}}-h_{\rm st},\qquad
\xi\equiv\sqrt{1+\frac{1}{h^2_{\rm st}}}-\frac{1}{h_{\rm st}},\qquad\eta=\sqrt{1+h_{\rm st}^2},\qquad\zeta=\frac{h_{\rm st}}{\xi}.
\end{equation}

It may be readily checked, that
\begin{equation}
h_{\rm st}=\frac{1-\kappa^2}{2\kappa},\qquad\xi=\frac{1-\kappa}{1+\kappa},\qquad\eta=\frac{1+\kappa^2}{2\kappa},\qquad\zeta=\frac{(1+\kappa)^2}{2\kappa}.
\end{equation}

As it is shown in the Appendix,
\begin{equation}\label{RHellip}
{\cal R}(\lambda,h_{\rm st})=I_{16}+4{\cal H}\lambda+o(\lambda).
\end{equation}

\section{Verification of the result}

In the two special cases,
\begin{equation}\label{trigon0}
h_{\rm st}=0\quad\Longrightarrow\quad\alpha(\lambda)=\tan{\lambda},\qquad\gamma(\lambda)=0,\qquad\theta(\lambda)=\frac{1}{\cos{\lambda}},
\end{equation}
and
\begin{equation}\label{trigonI}
h_{\rm st}=i\quad\Longrightarrow\quad\alpha(\lambda)=\tanh{\lambda},\qquad\gamma(\lambda)=i\tanh{\lambda},\qquad\theta(\lambda)=1,
\end{equation}
the verification of \eqref{YB} may be done in straightforward (but cumbersome) way. In order to economize the reader's time, we have presented the corresponding MAPLE 7 program.

In the general, elliptic, case the complete rigorous verification of \eqref{YB} seems to be far from the elementary theory of elliptic functions \cite{8}, so that the corresponding machinery is unclear for the author.
Nevertheless a strong evidence for implementation of
\eqref{YB} may be obtained by its verifications up to the orders $o(\lambda^j\mu^{m-j})$ with $m=0,\dots,M$ and rather big $M$. The corresponding MAPLE 7 programs are also presented in the Appendix. As a supplementary result we present (up to the orders $o(\lambda^j\mu^{m-j})$) the proofs of the following relations between $\alpha(\lambda)$, $\gamma(\lambda)$ and $\theta(\lambda)$
\begin{eqnarray}\label{rel1}
&&\gamma(1+\alpha^2)=h_{\rm st}\alpha(1-\gamma^2),\\\label{rel2}
&&\gamma\theta^2=h_{\rm st}\alpha,\\\label{rel3}
&&\frac{d\alpha}{d\lambda}=1+\alpha^2+2h_{\rm st}\alpha\gamma,\\\label{rel4}
&&\frac{d\gamma}{d\lambda}=h_{\rm st}(1-\gamma^2)-2\alpha\gamma.
\end{eqnarray}
The rigorous proofs for these formulas also are unclear for the author. The values of the governing parameter $M$ in the files "YBellip" and "relations" are adapted to the characteristics of the authors personal computer. The reader may increase them.

\appendix
\renewcommand{\theequation}{\thesection.\arabic{equation}}

\section{The verification MAPLE 7 programs}
\setcounter{equation}{0}

\subsection{The file "YBellip"}

Here we verify Eq. \eqref{YB} up to the order $M$ as it was discussed in the Sect. 3. The notations $\tt R1$, $\tt R2$, and $\tt R3$ mean
$R(\lambda-\mu)$, $R(\lambda)$, and $R(\mu)$ (the matrix $\tt R4$ will be used in the file "HR"). All of them are initialized in the file "dataR". The procedures {\tt tz} and {\tt extract} are given in the file "procedures". For two matrices $A$ and $B$ the output of ${\tt tz}(A,B)$ is the tensor product $A\otimes B$ (unfortunately the author did not found the analogous procedure in the MAPLE 7 library). The procedure {\tt extract} extracts polynomials of the given degree. For example
being applied to $p=1+x+2xy+3x^2+y^3$ the $\tt extract(p,x,y,2)$ will give $1+x+2xy+3x^2$. The correct output for "YB" should be $r=0$.

\begin{eqnarray}
&&{\tt with(linalg):}\nonumber\\
&&{\tt M:=8;}\nonumber\\
&&{\tt read(procedures):\,\, read(Jacobi):\,\, read(dataR):}\nonumber\\
&&{\tt I4:=array(identity,1..4,1..4):}\nonumber\\
&&{\tt Z:=evalm(tz(R1,I4)\&*tz(I4,R2)\&*tz(R3,I4)}\nonumber\\
&&{\tt-tz(I4,R3)\&*tz(R2,I4)\&*tz(I4,R1)):}\nonumber\\
&&{\tt for\,\, i\,\, to\,\, 64\,\, do\,\, for\,\, j\,\, to\,\, 64\,\, do}\nonumber\\
&&{\tt Z[i,j]:=extract(Z[i,j],lambda,mu,M):}\nonumber\\
&&{\tt od\,\,od:}\nonumber\\
&&{\tt r:=rank(Z);}
\end{eqnarray}

\subsection{The file "YBtrigon"}

Here the exact verifications of \eqref{YB} are given for the cases \eqref{trigon0} and \eqref{trigonI}. All notations are the same as in the previous Section. The file "Jacobi" is replaced by "trigon", presented in the Sect. A.7 in the form adapted to verification of \eqref{YB} for the $R$-matrix \eqref{trigon0}. In order to verify the case \eqref{trigonI} one has to make (in "trigon") the following replacements:
\begin{eqnarray}
&&({\tt h:=0;\rightarrow\#\tt h:=0;}),\qquad
({\tt\# h:=I;\rightarrow h:=I;)},\nonumber\\
&&({\tt c[i]:=0:\dots\rightarrow\#\tt c[i]:=0:\dots)},\qquad
({\#\tt b[i]:=1:\dots\rightarrow\tt b[i]:=1:\dots)}.
\end{eqnarray}
The correct output should be $r=0$.
\begin{eqnarray}
&&{\tt with(linalg):}\nonumber\\
&&{\tt read(procedures):\,\, read(trigon):\,\, read(dataR):}\nonumber\\
&&{\tt I4:=array(identity,1..4,1..4):}\nonumber\\
&&{\tt Z:=evalm(tz(R1,I4)\&*tz(I4,R2)\&*tz(R3,I4)}\nonumber\\
&&{\tt-tz(I4,R3)\&*tz(R2,I4)\&*tz(I4,R1)):}\nonumber\\
&&{\tt r:=rank(Z);}
\end{eqnarray}

\subsection{The file "relations"}

Here we give the verifications (up to the order $M$) of Eqs. \eqref{rel1}-\eqref{rel4}. The notation $\tt h$ means $h_{\rm st}$. The notations $\tt aa$, $\tt cc$, and $\tt bb$ mean $\alpha(x)$, $\gamma(x)$ and $\theta(x)$ expanded up to the order $o(x^M)$ (the corresponding constructions are presented in "Jacobi"). Such expansion can be worked out with the use of the procedure $\tt tp(f,m)$ (given in the file "procedures"). For example being applied to ${\tt f}={\rm e}^x$ at ${\tt m}=2$ it results in $1+x+x^2/2$. The correct output for the file "relations" should be $X=Y=Z=T=0$.

\begin{eqnarray}
&&{\tt with(linalg):}\nonumber\\
&&{\tt M:=30;}\nonumber\\
&&{\tt read(procedures):\,\, read(Jacobi):}\nonumber\\
&&{\tt X:=simplify(tp(c[4]*(1+a[4]^{\wedge}2)-h*a[4]*(1-c[4]^{\wedge}2),M));}\nonumber\\
&&{\tt Y:=simplify(tp(h*a[4]-c[4]*b[4]^{\wedge}2,M));}\nonumber\\
&&{\tt Z:=simplify(diff(a[4],x)-1-tp(a[4]^{\wedge}2+2*h*a[4]*c[4],M-1));}\nonumber\\
&&{\tt T:=simplify(diff(c[4],x)-h+tp(h*c[4]^{\wedge}2+2*a[4]*c[4],M-1));}
\end{eqnarray}

\subsection{The file "HR"}

Here we give the proof of the Eq. \eqref{RHellip}. The matrices {\tt HP}, {\tt HM}, and {\tt H} correspond to $H^+$, $H^-$ and ${\cal H}$. The correct output should be $r=0$.

\begin{eqnarray}
&&{\tt with(linalg):}\nonumber\\
&&{\tt M:=1;}\nonumber\\
&&{\tt read(procedures):}\nonumber\\
&&{\tt read(Jacobi):}\nonumber\\
&&{\tt read(dataR):}\nonumber\\
&&{\tt I2:=array(identity,1..2,1..2):}\nonumber\\
&&{\tt I4:=array(identity,1..4,1..4):}\nonumber\\
&&{\tt I16:=array(identity,1..16,1..16):}\nonumber\\
&&{\tt s3:=matrix(2,2,[1/2,0,0,-1/2]):}\nonumber\\
&&{\tt sp:=matrix(2,2,[0,1,0,0]):}\nonumber\\
&&{\tt sm:=matrix(2,2,[0,0,1,0]):}\nonumber\\
&&{\tt HP:=evalm((stz(sp,sm)+h*atz(I2,s3))/2):}\nonumber\\
&&{\tt HM:=evalm((stz(sp,sm)-h*atz(I2,s3))/2):}\nonumber\\
&&{\tt H:=evalm(stz(HM,I4)/2+tz(tz(I2,HP),I2)):}\nonumber\\
&&{\tt X:=evalm(R4-I16-4*x*H):}\nonumber\\
&&{\tt for\,\, i\,\, to\,\, 16\,\, do\,\, for\,\, j\,\, to\,\, 16\,\, do}\nonumber\\
&&{\tt X[i,j]:=tp(X[i,j],M)}\nonumber\\
&&{\tt od\,\, od:}\nonumber\\
&&{\tt r:=rank(X);}
\end{eqnarray}

\subsection{The file "procedures"}

\begin{eqnarray}
&&{\tt tz:=proc(A,B)\, local\, a,b,c,C,i,j,k,l:}\nonumber\\
&&{\tt a:=rowdim(A):\,\, b:=rowdim(B):\,\, c:=a*b:}\nonumber\\
&&{\tt C:=array(1..c,1..c):}\nonumber\\
&&{\tt for\,\, i\,\, to\,\, a\,\, do\,\, for\,\, j\,\, to\,\, a\,\, do\,\, for\,\, k\,\, to\,\, b\,\, do\,\, for\,\, l\,\, to\,\, b\,\, do}\nonumber\\
&&{\tt C[b*(i-1)+k,b*(j-1)+l]:=A[i,j]*B[k,l]:}\nonumber\\
&&{\tt od\,\, od\,\, od\,\, od:}\nonumber\\
&&{\tt tz(A,B):=evalm(C):}\nonumber\\
&&{\tt end:}\nonumber\\
&&\nonumber\\
&&{\tt stz:=proc(A,B):}\nonumber\\
&&{\tt stz(A,B):=evalm(tz(A,B)+tz(B,A)):}\nonumber\\
&&{\tt end:}\nonumber\\
&&\nonumber\\
&&{\tt atz:=proc(A,B):}\nonumber\\
&&{\tt atz(A,B):=evalm(tz(A,B)-tz(B,A)):}\nonumber\\
&&{\tt end:}\nonumber\\
&&\nonumber\\
&&{\tt tp:=proc(w,m):}\nonumber\\
&&{\tt tp(w,m):=expand(convert(taylor(w,x,m+1),polynom)):}\nonumber\\
&&{\tt end:}\nonumber\\
&&\nonumber\\
&&{\tt extract:=proc(p,x,y,m)\,\, local\,\, i,j,k,l:}\nonumber\\
&&{\tt extract(p,x,y,m):=sum('sum('factor(coeff(coeff(p,x,k),y,l-k)})\nonumber\\
&&{\tt *x^{\wedge}k*y^{\wedge}(l-k)','k'=0..l)','l'=0..m):}\nonumber\\
&&{\tt end:}
\end{eqnarray}

\subsection{The file "Jacobi"}

Here {\tt a[4]}, {\tt b[4]}, and {\tt c[4]} are the power expansions for $\alpha(x)$, $\theta(x)$, and $\gamma(x)$ up to the order $o(x^M)$.
The procedure {\tt subst} turns $x$ into $\lambda-\mu$, $\lambda$, and $\mu$.

\begin{eqnarray}
&&{\tt z:=array(1..3):}\nonumber\\
&&{\tt z[1]:=lambda-mu:\,\, z[2]:=lambda:\,\, z[3]:=mu:}\nonumber\\
&&{\tt xi:=(1-kappa)/(1+kappa):\,\, eta:=(1+kappa^{\wedge}2)/2/kappa:}\nonumber\\
&&{\tt zeta:=(1+kappa)^{\wedge}2/2/kappa:\,\, h:=(1-kappa^{\wedge}2)/2/kappa:}\nonumber\\
&&{\tt a:=array(1..4):\,\, b:=array(1..4):\,\, c:=array(1..4):}\nonumber\\
&&{\tt a[4]:=tp(I*kappa*JacobiSN(x/I/kappa,kappa^{\wedge}2),M):}\nonumber\\
&&{\tt b[4]:=tp(JacobiCN(I*eta*x,1/eta),M):}\nonumber\\
&&{\tt c[4]:=tp(xi*JacobiSN(zeta*x,xi^{\wedge}2),M):}\nonumber\\
&&{\tt for\,\, i\,\, to\,\, 3\,\, do}\nonumber\\
&&{\tt a[i]:=subst(a[4],x,z[i],M):\,\, b[i]:=subst(b[4],x,z[i],M):}\nonumber\\
&&{\tt c[i]:=subst(c[4],x,z[i],M):}\nonumber\\
&&{\tt od:}\nonumber\\
\end{eqnarray}

\subsection{The file "trigon"}

\begin{eqnarray}
&&{\tt z:=array(1..3):}\nonumber\\
&&{\tt h=0;}\nonumber\\
&&{\#\tt h:=I;}\nonumber\\
&&{\tt z[1]:=x/y: z[2]:=x: z[3]:=y:}\nonumber\\
&&{\tt a:=array(1..4):\,\, b:=array(1..4):\,\, c:=array(1..4):}\nonumber\\
&&{\tt for\,\, i\,\, to\,\, 3\,\, do}\nonumber\\
&&{\tt c[i]:=0: a[i]:=(z[i]-1/z[i])/(z[i]+1/z[i])/I: b[i]:=2/(z[i]+1/z[i]):}\nonumber\\
&&{\#\tt b[i]:=1: a[i]:=(z[i]-1/z[i])/(z[i]+1/z[i]): c[i]:=I*a[i]:}\nonumber\\
&&{\tt od:}\nonumber\\
&&{\tt a[4]:=0: b[4]:=0: c[4]:=0:}
\end{eqnarray}

\subsection{The file "dataR"}

\begin{eqnarray}
&&{\tt for\,\, i\,\, to\,\, 4\,\, do}\nonumber\\
&&{\tt A||i:=array(symmetric,1..16,1..16):}\nonumber\\
&&{\tt B||i:=array(symmetric,1..16,1..16):}\nonumber\\
&&{\tt for\,\, j\,\, to\,\, 16\,\, do\,\, for\,\, k\,\, to\,\, 16\,\, do}\nonumber\\
&&{\tt A||i[j,k]:=0:\,\, B||i[j,k]:=0:}\nonumber\\
&&{\tt od\,\, od:}\nonumber\\
&&{\tt A||i[1,1]:=1:\,\, A||i[2,2]:=b[i]:\,\, A||i[8,8]:=A||i[2,2]:}\nonumber\\
&&{\tt A||i[3,3]:=(1+2*a[i]^{\wedge}2+4*h*a[i]*c[i])*b[i]:\,\, A||i[5,5]:=A||i[3,3]:}\nonumber\\
&&{\tt A||i[4,4]:=(1+a[i]^{\wedge}2)*(1+a[i]^{\wedge}2+2*h*a[i]*c[i]):}\nonumber\\
&&{\tt A||i[6,6]:=1+(3+8*h^{\wedge}2)*a[i]^{\wedge}2+a[i]^{\wedge}4:}\nonumber\\
&&{\tt A||i[7,7]:=(1+a[i]^{\wedge}2):\,\,
A||i[2,3]:=a[i]*b[i]:\,\, A||i[5,9]:=A||i[2,3]:}\nonumber\\
&&{\tt A||i[2,5]:=a[i]^{\wedge}2:\,\,A||i[3,9]:=A||i[2,5]:\,\,A||i[7,10]:=A||i[2,5]:}\nonumber\\
&&{\tt A||i[4,6]:=a[i]*(2+a[i]^{\wedge}2+4*h*a[i]*c[i])*b[i]:}\nonumber\\
&&{\tt A||i[4,7]:=a[i]^{\wedge}2*b[i]:\,\, A||i[4,10]=A||i[4,7]:}\nonumber\\
&&{\tt A||i[4,11]:=a[i]^{\wedge}3*b[i]:\,\, A||i[4,13]:=a[i]^{\wedge}4:}\nonumber\\
&&{\tt A||i[6,7]:=a[i]*(1+a[i]^{\wedge}2+2*h*a[i]*c[i]):\,\, A||i[6,10]=A||i[6,7]:}\nonumber\\
&&{\tt A||i[3,5]=2*A||i[6,7]:\,\,A||i[6,11]:=a[i]^{\wedge}2*(1+a[i]^{\wedge}2):}\nonumber\\
&&{\tt B||i[2,2]:=c[i]*b[i]:\,\, B||i[8,8]:=B||i[2,2]:}\nonumber\\
&&{\tt B||i[5,5]:=c[i]*(3+2*a[i]^{\wedge}2+4*h*a[i]*c[i])*b[i]:}\nonumber\\
&&{\tt B||i[3,3]=-B||i[5,5]:\,\, B||i[4,4]=2*h*a[i]*(a[i]^{\wedge}2-1):}\nonumber\\
&&{\tt B||i[4,6]=a[i]*c[i]*(2+3*a[i]^{\wedge}2+4*h*a[i]*c[i])*b[i]:}\nonumber\\
&&{\tt B||i[4,7]=a[i]^{\wedge}2*c[i]*b[i]:\,\, B||i[4,10]:=B||i[4,7]:}\nonumber\\
&&{\tt B||i[5,9]=a[i]*c[i]*b[i]:\,\, B||i[2,3]:=-B||i[5,9]:}\nonumber\\
&&{\tt B||i[6,7]=2*h*a[i]^{\wedge}2:\,\, B||i[6,10]:=B||i[6,7]}\nonumber\\
&&{\tt B||i[4,11]=-a[i]^{\wedge}3*c[i]*b[i]:\,\, B||i[6,6]=4*h*A||i[6,7]:}\nonumber\\
&&{\tt for\,\, j\,\, to\,\, 8\,\, do\,\, for\,\, k\,\, to\,\, 16\,\, do}\nonumber\\
&&{\tt A||i[17-j,17-k]:=A||i[j,k]:\,\, B||i[17-j,17-k]:=-B||i[j,k]:}\nonumber\\
&&{\tt od\,\, od:}\nonumber\\
&&{\tt R||i:=evalm(A||i+B||i):}\nonumber\\
&&{\tt od:}
\end{eqnarray}

\end{document}